\documentclass[%
11pt,
reprint,
onecolumn,
tightenlines,
superscriptaddress,
preprintnumbers,
nofootinbib,
amsmath,amssymb,amsthm,
physrev,
eqsecnum,tikz,
]{revtex4-2}

\usepackage{isomath}
\usepackage{amsmath,amsthm}
\usepackage{amsbsy}
\usepackage{amssymb}
\usepackage{amscd}
\usepackage{amsfonts}
\usepackage{stmaryrd}
\usepackage{siunitx}
\usepackage{euscript}
\usepackage[utf8]{inputenc}
\usepackage[T1]{fontenc}
\usepackage{newtxtext} 
\everymath{\displaystyle}
\usepackage{exscale}
\usepackage{microtype}
\usepackage{hyperref}
\usepackage{booktabs}
\usepackage{algorithm}
\usepackage{algpseudocode}

\usepackage{graphicx}
\usepackage{boxedminipage}
\usepackage{calc}
\usepackage[dvipsnames]{xcolor}

\usepackage{setspace}
\usepackage{enumitem}
\setitemize{noitemsep,topsep=0pt,parsep=0pt,partopsep=0pt}
\setenumerate{noitemsep,topsep=0pt,parsep=0pt,partopsep=0pt}
\setdescription{noitemsep,topsep=0pt,parsep=0pt,partopsep=0pt}

\usepackage{orcidlink}
\usepackage{siunitx}
\usepackage[small]{titlesec}

\titlespacing*{\section}{0pt}{12pt plus 4pt minus 2pt}{2pt plus 2pt minus 2pt}
\titlespacing*{\subsection}{0pt}{12pt plus 4pt minus 2pt}{2pt plus 2pt minus 2pt}
\titlespacing*{\subsubsection}{0pt}{12pt plus 4pt minus 2pt}{2pt plus 2pt minus 2pt}
\titlespacing*{\paragraph}{0pt}{12pt plus 4pt minus 2pt}{2pt plus 2pt minus 2pt}

\makeatletter

    \renewcommand*{\p@subsection}{}
    
    \renewcommand*{\p@subsubsection}{}
\makeatother

\usepackage{isomath}
\usepackage{amsmath}
\usepackage{amssymb}
\usepackage{amscd}
\usepackage{amsfonts}
\usepackage{upgreek}

\usepackage{subfig}
\usepackage{graphicx}

\newcommand{\R}{\mathbb R}

\newcommand{\eps}{{\varepsilon}}

\newcommand{\half}{\tfrac{1}{2}}

\theoremstyle{definition}
\newtheorem{definition}{Definition}[section]
\AtEndEnvironment{definition}{\null\hfill\qedsymbol}

\AtEndEnvironment{remark}{\null\hfill\qedsymbol}

\AtEndEnvironment{example}{\null\hfill\qedsymbol}

\AtEndEnvironment{assumption}{\null\hfill\qedsymbol}

\newcommand{\bfSigma}{\mathbold {\Sigma}}

\DeclareMathOperator{\trace}{tr}
\DeclareMathOperator{\range}{range}
\DeclareMathOperator{\codim}{codim}

\newcommand{\const}{\mathrm{const.}}

\newcommand{\parderiv}[2]{\frac{\partial #1}{\partial #2}}
\newcommand{\dm}{\ \mathrm{d}}

\newcommand{\bfe}{{\mathbold e}}

\newcommand{\bfn}{{\mathbold n}}

\newcommand{\bft}{{\mathbold t}}

\newcommand{\bfx}{{\mathbold x}}

\newcommand{\bfE}{{\mathbold E}}
\newcommand{\bfF}{{\mathbold F}}

\newcommand{\bfP}{{\mathbold P}}

\newcommand{\bfT}{{\mathbold T}}
\newcommand{\bfU}{{\mathbold U}}

\newcommand{\hE}{\hat{E}}
\newcommand{\hpsi}{\hat{\psi}}
\newcommand{\hT}{\hat{T}}
\newcommand{\that}{\hat{t}}
\newcommand{\thatb}{\hat{t}_{\mathrm{bif}}}

\newcommand{\la}{\lambda}
\newcommand{\lab}{\la_{\mathrm{bif}}}
\newcommand{\Eext}{E^{\mathrm{ext}}}
\newcommand{\bfEext}{\mathbold E^{\mathrm{ext}}}

\newcommand{\ga}{\gamma}

\newcommand{\RN}[1]{\textup{\uppercase\expandafter{\romannumeral#1}}}

\newcommand{\I}{\RN{1}}
\newcommand{\II}{\RN{2}}

\begin{document}

\preprint{To appear in Journal of Applied Mechanics (DOI: \href{https://doi.org/10.1115/1.4068630}{10.1115/1.4068630})}

\title{Soft Electromechanical Elastomers Impervious to Instability}

\author{Daniel Katusele}
    \email{dkatusel@andrew.cmu.edu}
    \affiliation{Department of Civil and Environmental Engineering, Carnegie Mellon University}

\author{Carmel Majidi}
     \affiliation{Department of Mechanical Engineering, Carnegie Mellon University}

\author{Kaushik Dayal \orcidlink{0000-0002-0516-3066}}
    \affiliation{Department of Civil and Environmental Engineering, Carnegie Mellon University}
    \affiliation{Center for Nonlinear Analysis, Department of Mathematical Sciences, Carnegie Mellon University}
    \affiliation{Department of Mechanical Engineering, Carnegie Mellon University}

\author{Pradeep Sharma}
    \affiliation{Department of Mechanical Engineering, University of Houston}

\begin{abstract}
    Soft dielectric elastomers that can exhibit extremely large deformations under the action of an electric field enable applications such as soft robotics, biomedical devices, energy harvesting among others. A key impediment in the use of dielectric elastomers is failure through instability mechanisms or dielectric breakdown. In this work, using a group-theory based approach, we provide a closed-form solution to the bifurcation problem of a paradigmatical elastomer actuator and discover an interesting result: at a critical electric field, the elastomer becomes impervious to Treloar-Kearsley instability. 
    This limit is reached \emph{prior} to the typical dielectric breakdown threshold.
    Our results thus establish a regime of electrical and mechanical loads where the dielectric elastomer is invulnerable to all common failure modes. 
\end{abstract}

\maketitle


\section{Introduction}

Soft electromechanical elastomers, also known as Dielectric Elastomers (DE), wherein mechanical deformations can be driven by electrical stimulus for actuation, present tremendous potential as transducers for soft and biologically-inspired robots, biomedical devices, energy harvesting, among other applications \cite{carpi2011bioinspired, rus2015design, kofod2007energy, pelrine2000highs,pelrine2000highf, bauer201425th, yang2017avoiding, koh2009maximal, bar2004electroactive, bartlett2016stretchable,grasinger2021flexoelectricity,grasinger2021architected,Zhang2024multielectrode, Li2023Bioinspired,chen2019controlled, Maziz2017Knitting,acome2018hydraulically}. 
The actuation mechanism in DE is typically achieved through a capacitor-like design where a dielectric elastomer film is sandwiched between two compliant electrode: upon application of a voltage difference across the electrodes, the electrostatic (Coulombic) force between the electrodes due to the electrical charges compresses the DE in the thickness direction causing --- through the Poisson effect --- the DE to expand in the lateral direction \cite{pelrine2000highstr, wissler2007mechanical, kollosche2012complex}.

However, because the Poisson effect is typically fairly small, DEs typically require high voltages to induce a usable deformation.
Consequently, the high fields drive electromechanical failures such as pull-in instability, electrical breakdown, and buckling instability \cite{zhao2014harnessing, bense2017buckling, lu2020mechanics} which limit the performance. 
To increase the deformation and delay/suppress instabilities, mechanical prestretch and prestress are often introduced before applying voltage \cite{li2011effect, kofod2008static, yang2017avoiding, huang2012giant}. 
Other methods to improve the performance of DE include the introduction of mechanical constraints \cite{zhao2014harnessing, zhang2011mechanical}, using dielectric films without electrodes \cite{keplinger2010rontgen}, designing multilayer DE actuators \cite{Shi2022}  or harnessing instabilities for improved functionality \cite{huang2012giant, katusele2025exploiting}.
A compelling argument for the use of soft dielectric elastomers (DEs) is their ability to sustain large mechanical deformation under the action of an electrical field. Such a feature is a necessity for applications like biomimetic robotics or devices such as an adaptive eye lens. The facile deformability and the complex nonlinear material behavior that underlies its mechanics, also gives rise to instabilities such as wrinkling, creasing, pull-in instability, Treloar-Kearsley instability and others \cite{dorfmann2019instabilities,chen2021interplay, yang2023tutorial}. While instabilities may be exploited to achieve interesting actuation designs\cite{Chi2022,shao2018bioinspired}, they are more typically the failure modes for dielectric elastomer based devices.

We briefly highlight the two instability mechanisms relevant to our work. For the capacitive thin film elastomer design, its thinning progressively increases with increasing electrical field. Once the film reaches a critical value, thinning increases suddenly (to be understood as a bifurcation) leading to electrical breakdown. The Treloar-Kearsley (T-K) instability is based on a symmetric deformation bifurcating into symmetry-breaking configuration beyond a critical applied stimulus e.g. square to rectangular or circular to elliptical \cite{batra2005treloar,steigmann2007simple,treloar1948stresses,kearsley1986asymmetric,wang2022axisymmetric,li2020kearsley,wineman2024treloar,Fu2023,Yu2025}.

In this letter we address the following question: is it possible to engineer a dielectric elastomer to be completely immune to instability? 
We first obtain a closed-form solution to the bifurcation analysis of a paradigmatic circular dielectric elastomer thin film subject to both electrical and mechanical loading.  
Building on this solution, we find that an applied electric field of a certain critical strength completely suppresses the T-K instability.

\section{Model Formulation: Variational Principle and Field Equations}
Consider a DE specimen occupying the domain $\Omega_0$ and boundary $\partial \Omega_0$ in the reference configuration, and $\Omega$ with boundary $\partial \Omega$ in the deformed configuration. 
The material points in the reference and deformed configurations are denoted by $\bfx_0$ and $\bfx$ respectively, and the deformation gradient by $\bfF(\bfx_0) = \nabla_0 \bfx$ consistent with the deformation map $\bfx = \bfx(\bfx_0)$; we further define $J = \det(\bfF) > 0$ as the Jacobian. 
The electric potential $\phi(\bfx)$ and electric field $\bfE(\bfx)$ are related by $\bfE = - \nabla \phi$.
The polarization field in the material is denoted as $\bfP(\bfx)$. 

The total free energy of the system is formulated \cite{liu2013energy} as 
\begin{equation} 
    \psi[\bfx, \bfP] 
    = 
    \int_{\Omega_0} W(\bfF, \bfP) + \frac{\eps_0}{2} \int_{\Omega} |\bfE|^2 - \int_{\partial \Omega_{t_0}} \bft_0 \cdot \bfx + \int_{\partial \Omega} \phi (\eps_0 \bfE + \bfP) \cdot \bfn \label{Elr_Engy}
\end{equation}
where $W$ is the free energy density per unit referential volume; the second term is the electrostatic field energy, noting that $\eps_0$ is the permittivity of free space; and the last two terms are the contributions from the  mechanical and electrical boundary conditions respectively. 
$\partial \Omega_{t_0}$ is the part of the boundary $\partial \Omega_0$ where the traction is specified with $\bfn$ the outward normal. 
The electric field in \eqref{Elr_Engy} is computed by solving the electrostatic equation :
\begin{equation}  
    \nabla \cdot (\eps_0 \bfE + \bfP) = -\eps_0 \nabla^2 \phi + \nabla \cdot \bfP = 0 \quad \text{in } \Omega \label{Maxwell}
\end{equation}
subject to the boundary conditions that $\phi$ is specified at the electrodes and $\left(\eps_0 \bfE + \bfP\right) \cdot \bfn = 0$ on the portion of the boundary where there are no free charges.
We note that this is an approximation that neglects the external electric fields outside the specimen \cite{yang2011completely,jha2023discrete}.



The pullbacks to the reference configuration of $\bfE(\bfx), \bfP(\bfx)$ are defined following \cite{marshall2014atomistic,jha2023atomic} to be:
\begin{equation} 
    \bfE_0 = \bfF^T \bfE, \quad \bfP_0 = J \bfP\nonumber 
\end{equation}
and the pullback for the electric potential to be $\phi_0(\bfx_0) = \phi(\bfx(\bfx_0))$. 

The dielectric elastomer is assumed to be incompressible, which requires that $J = 1$, and is imposed by introducing a Lagrange multiplier $p(\bfx_0)$.
The Lagrangian for an incompressible material, written in terms of the pullbacks, has the expression:
\begin{equation}
    \Psi[\bfx, \bfP_0] = \int_{\Omega_0} W(\bfF, \bfP_0) + \frac{\eps_0}{2} \int_{\Omega_0} J |\bfF^{-T} \bfE_0|^2 - \int_{\partial \Omega_{t_0}} \bft_0 \cdot \bfx  + \int_{\partial \Omega_0} \phi_0 J \bfF^{-1}( \eps_0 \bfF^{-T} \bfE_0 + \bfP_0)  \cdot \bfn_0 - \int_{\Omega_0} p (J-1) \label{Lgr_Engy}
\end{equation}  
where $\bfn_0$ is the outward normal to $\partial \Omega_0$. 

Setting the functional derivative of $\Psi$ with respect to $\bfx(\bfx_0)$ to $0$, with the constraint \eqref{Maxwell}, we obtain the following equations that represent mechanical equilibrium in the bulk and the boundary conditions:
\begin{subequations}
\begin{align} 
    \nabla_0 \cdot \left( \frac{\partial W}{\partial \bfF} + \bfSigma_0 - p J \bfF^{-T}\right) = \mathbf{0}  & \quad \text{on } \Omega_0 \label{equil1} \\
    \left( \frac{\partial W}{\partial \bfF} + \bfSigma_0 - p J \bfF^{-T}\right) \bfn_0 = \bft_0 & \quad \text{on } \partial \Omega_{t_0} \label{equil2} \\
    \left( \frac{\partial W}{\partial \bfF} + \bfSigma_0 - p J \bfF^{-T}\right) \bfn_0 = \mathbf{0} & \quad \text{on } \partial \Omega_0 \setminus \partial \Omega_{t_0} \label{equil3}
\end{align}
\label{eqn:equil}
\end{subequations}
We have defined $\bfSigma_0 := \bfE_0 \otimes J \bfF^{-1}( \eps_0 \bfF^{-T} \bfE_0 + \bfP_0) - \frac{\eps_0 J}{2} |\bfE_0|^2 \bfF^{-T}$ as the Piola-Maxwell stress tensor, and $\bfT := \frac{\partial W}{\partial \bfF} + \bfSigma_0 - p J \bfF^{-T}$ as the total Piola-Kirchhoff stress tensor.
The PDE and BCs in \eqref{eqn:equil} define the boundary value problem (BVP) that must be solved for the equilibrium configuration.

Similarly, setting the functional derivative of $\Psi$ with respect to $\bfP_0(\bfx_0)$ to $0$ gives the usual local relation between the electric field and polarization density at a point $-\parderiv{W}{\bfP_0} = \bfE$ \cite{darbaniyan2019designing}.
\subsection{Material Model}
We assume that the energy density $W(\bfF, \bfP_0)$ is additively composed of a mechanical strain energy density $W^{\text{m}}(\bfF)$ and an electromechanical energy density $W^{\text{el}}(\bfF, \bfP_0)$. 

For the mechanical term, we use an incompressible, isotropic, hyperelastic Mooney-Rivlin model \cite{rivlin1948large}, that can be connected to statistical mechanics and network elasticity \cite{khandagale2023statistical,grasinger2023polymer}, with the form:
\begin{equation} 
    W^{\text{m}}(\bfF) = \frac{\mu}{2}\left[\left(\I_1 - 3 \right) + \ga \left( \I_2 - 3 \right) \right]  \label{strain-engy}
\end{equation}
where $\mu$ and $\ga$ are material parameters, and $\I_1  = \trace (\bfF^T \bfF)$ and $\I_2 = \half \left(\trace (\bfF^T \bfF)^2 - \trace ((\bfF^T \bfF)^2)\right)$ are the invariants of $\bfF$. 
All quantities will be non-dimensionalized with respect to $\mu$.

For the electromechanical term, we use a linear isotropic dielectric with $\bfP = \eps_0 \chi \bfE$, where $\chi$ is the scalar dielectric susceptibility. 
In \cite{grasinger2021nonlinear}, it was shown that the dielectric susceptibility derived from statistical mechanics is an anisotropic function of the deformation; however, for simplicity, we assume that $\chi$ is isotropic and independent of deformation. 
Defining $\eps := \eps_0(1+ \chi)$ to be the permittivity, we write the electromechanical energy as \cite{liu2014energy}:
\begin{align}
    W^{\text{el}} = \frac{1}{2J} \bfP_0 \cdot (\eps - \eps_0)^{-1} \bfP_0.
    \label{eq:elec_energy}
\end{align}
We note that though $\eps$ is independent of deformation, $W^{\text{el}}$ involves the deformation through the presence of $J=\det \bfF$ and because $\bfP_0$ depends on $\bfF$ through the pullback relation.
\section{Simplification of the Field Equations to an Algebraic System} \label{bifurc}

The necessary conditions for the onset of the symmetry-breaking instability is determined through a linear bifurcation analysis on a DE specimen subject to both mechanical loads and electrical stimuli. 
We consider a disk-shaped specimen, and the deformation and electric field are both assumed to be homogeneous under the applied loads.
This enables us to simplify our analysis for TK and pull-in instabilities, but restricts it to situations without buckling instabilities.

The Cartesian coordinates of material points in the reference configuration are of the form $\bfx_0 = (x_1, x_2, x_3) = (R \cos \theta, R \sin \theta, x_3)$, where $R$ is the radius of the disk and $\theta\in [0,2\pi)$.
The corresponding spatial position of these material points after deformation is of the form $\bfx = (F_{11} x_1 + F_{12} x_2, F_{21} x_1 + F_{22} x_2, F_{33} x_3)$, with 
\begin{equation} 
    \bfF 
    = 
    \begin{pmatrix}
        F_{11} & F_{12} & 0\\
        F_{21} & F_{22} & 0 \\
        0 & 0 & F_{33}
    \end{pmatrix} 
\end{equation}
The components $F_{13}, F_{23}, F_{31}$ and $F_{32}$ are negligible because the specimen has a thickness that is small compared to its radius.

The mechanical load is applied uniformly on the entire lateral boundary by specifying the traction $\bfT \bfe_r = \bft_0$, with $\bft_0 = t_0 \bfe_r$ and $\bfe_r = (\cos{\theta}, \sin \theta, 0)$.
The top and bottom faces are traction free, i.e $\bfT \bfe_3 = \bf0$. 
The term $\bft_0 \cdot \bfx$ in \eqref{Lgr_Engy} evaluates to $Rt_0((F_{11}x_1 + F_{12}x_2)\cos \theta + (F_{21}x_1 + F_{22}x_2)\sin \theta))$. 
The energy contribution due to the applied traction can now be written as
\begin{align}
    \int_{\partial \Omega_{t_0}} \bft_0 \cdot \bfx = RHt_0 \int_0^{2 \pi} (F_{11}\cos^2 \theta + F_{22} \sin^2 \theta + (F_{12} + F_{21}) \cos \theta \sin \theta)R \dm \theta = \pi R^2Ht_0 (F_{11} + F_{22})
\end{align}

The voltage boundary condition has the affine form $\phi = -\bfEext \cdot \bfx$ and is applied on the entire boundary $\partial\Omega$, where $\bfEext = (\Eext_1, \Eext_2, \Eext_3)$ is a constant vector with the physical interpretation of a uniform applied electric field \cite{grasinger2020statistical,grasinger2022statistical,khandagale2024statistical}.
The internal electric field that is generated by this boundary voltage is computed from \eqref{Maxwell} and evaluates to $\nabla \phi = -\bfEext$. 

The energy due to the electric field in \eqref{Lgr_Engy}, by using \eqref{eq:elec_energy} and under the assumption of a homogeneous deformation, simplifies to:
\begin{equation}
    \frac{\eps}{2} \int_{\Omega_0} J |\bfF^{-T} \bfEext|^2 + \int_{\partial \Omega_0} \phi_0 \bfEext  \cdot \bfn_0 = -\pi R^2 H\frac{\eps}{2} |\bfF^{-T} \bfEext|^2 
\end{equation}
where we have used the prescribed affine voltage boundary conditions and the divergence theorem.


Using all the simplified expressions from above, the mean free energy, in terms of components of $\bfF$ and $\bfE^{ext}$ can be written as:
\begin{equation}
\begin{split}
    \frac{1}{\pi R^2H}\psi(F_{11}, F_{12}, F_{21}, F_{22}, F_{33}) 
    = 
    & 
    \frac{\mu}{2}\left((\I - 3) + \ga( \II - 3)\right) - t_0(F_{11} + F_{22}) 
    - \frac{\eps}{2} (F_{22}\Eext_1 - F_{21}\Eext_2)^2F_{33}^2 
    \\
    & 
    - \frac{\eps}{2} (F_{11}\Eext_2 - F_{12}\Eext_1)^2F_{33}^2 - \frac{\eps}{2} \frac{(\Eext_3)^2}{F_{33}^2} - p \left(F_{11}F_{22}F_{33}-F_{12}F_{21}F_{33} -1\right)
\end{split}
 \label{comp_engy}
\end{equation}
This generalizes the expression from \cite{chen2021interplay} which was restricted to out-of-plane electric fields.


To further simplify our analysis, we will determine the condition for onset of the T-K instability in the presence of in-plane components of the electric field in terms of principal stretches  only, which are the eigenvalue of the stretch tensor $\bfU = \sqrt{\bfF^T \bfF}$. It is the case that prior to a critical value for bifurcation, the dielectric elastomer will stretch uniformly in its plane, i.e it will remain circular, thus there will be no shearing. Incompressibility of the dielectric requires that $\det(\bfF) = 1$; that is $\la_1 \la_2 \la_3 = 1$. The traction BC on the top and bottom, referring to \eqref{equil3}, indicates zero stress components in the out-of-plane direction (that is $T_3$); this eliminates $p$. We further set the electric field components $\Eext_1$ and $\Eext_2$ to zero given that the analysis herein focuses on the effect of $\Eext_3$ on the instabilities. It was shown in \cite{katusele2025exploiting} that both $\Eext_1$ and $\Eext_2$ had minimal effect on the T-K instability compared to $\Eext_3$. The mean free energy and in-plane principal stress components $\hT_1$ and $\hT_2$ as defined from \eqref{equil1}, given by the algebraic system of equations as functions of principal stretches $\la_1, \la_2$ and $\hE_3$ in the form
\begin{equation}
    \frac{1}{\pi R^2H}\hpsi(\la_1, \la_2, \hE_3) 
    = \frac{1}{2}\left(\la^2_1 + \la^2_2 + \frac{1}{\la^2_1 \la^2_2} - 3 + \ga( \la^{-2}_1 + \la^{-2}_2 + \la^2_1 \la^2_2 - 3)\right) - 
    \that_0(\la_1 + \la_2) 
     - \frac{1}{2} \hE^2_3 \la^2_1 \la^2_2
 \label{comp_engy_lambda}
\end{equation}
\begin{subequations}
\begin{align} \label{str1}
\hT_{1}(\la_1, \la_2, \hE_3) := - \hE_3^2 \la_1 \la_2^2 - \la_1^{-3} \la_2^{-2} + \la_1 + \ga ( \la_1^4 \la_2^2 - 1) \la_1^{-3} \\ \label{str2}
\hT_{2}(\la_1, \la_2, \hE_3) :=  - \hE_3^2 \la_1^2\la_2 - \la_1^{-2}\la_2^{-3} + \la_2 + \ga ( \la_1^2 \la_2^4 - 1)\la_2^{-3} 
\end{align} \end{subequations}
where $\hpsi(\la_1, \la_2, \hE_3) := \psi[F_{11}, F_{12}, F_{21}, F_{22}, F_{33}]/ \mu$, $\that_0 := \frac{t_0}{\mu}$, $\hT_i := \frac{T_i}{\mu}$, $i=1,2$; and $\hE_3 := \frac{\Eext_3}{\sqrt{\mu / \eps}}$

For subsequent analysis, the total nominal stress will be $\hT_i = \that_0$, where $i=1,2$ based on the boundary condition in \eqref{equil2}. Then, eliminating $t_o$ from \eqref{str1} and \eqref{str2}, we get the condition
\begin{align} \label{sol_bvp}
\left(\la_1 - \la_2 \right)\left(\left(\hE^2_3 - \ga\right) \la_1\la_2 +(1 + \ga (\la_1^2 + \la_1\la_2 + \la_2^2)) \la_1^{-3}\la_2^{-3} + 1 \right) = 0
\end{align}
In this absence of electric fields ($\hE_3 = 0$), the condition above matches Kearsley's theoretical analysis in predicting asymmetric deformation beyond a critical load \cite{kearsley1986asymmetric}.

\section{Background on Singularity Theory}

A bifurcation analysis of the dielectric employing singularity theory is presented below. This section is adapted from \cite{chen2001singularity} with references from \cite{golubitsky2012singularitiesI} and \cite{golubitsky2012singularitiesII}.
Singularity theory is a mathematical tool that seeks to reduce a singular function to a simple normal form from which the properties of the bifurcation solution can be determined from a finite number of derivatives of the singular function. 
We will be concerned with local bifurcation problems of the form 
\begin{equation}
    f(u, \kappa) = 0
    \label{eq:infinite_dim_equation}
\end{equation}
near a point $(u_0, \kappa_0)$, where $u$ represents state variables and $\kappa$ is the bifurcation parameter. Classically, $(u_0, \kappa_0)$ is called a bifurcation point if the number of solutions changes as $\kappa$ changes  in the neighborhood of $\kappa_0$. We consider the principal branch $u$ bifurcating near $u_0$ when $\kappa$ increases past $\kappa_0$. 

An illustrative example of a bifurcation is the elastica problem of buckling of a column subjected to compressive forces \cite{antman1995nonlinear}. 
The principal branch is the unbuckled state while the buckled state is the bifurcated branch. 
The T-K instability is another example of a bifurcation problem where a Mooney-Rivlin material stretched biaxially experience a symmetric stretch up to a bifurcation where the stretch becomes asymmetric leading to stable and unstable solutions \cite{kearsley1986asymmetric}. 
In this section will will explore the former to illustrate the singularity theory and the latter will be analysed in the subsequent sections.

The governing equation for the elastica problem is:
\begin{align}
    EIu''(s) + \kappa \sin u(s) = 0, \quad 0<s<l,
    \label{eq:elastica}
\end{align}
where $u$ is the angle between the undeformed rod and the tangent of the deformed rod, $s$ the material coordinate, $E$ the elastic modulus, $I$ the moment of inertia, $\kappa$ the compressive applied force, and $l$ the length of the rod.  $u$ is a state variable and $\kappa$ is the bifurcation parameter. The rod is hinged at its ends, with the boundary conditions:
\begin{align}
    u'(0) = u'(l) = 0.
    \label{eq:elastica_bc}
\end{align}
In linear beam theory, the assumption $|u(s)| \ll 1$ leads to a linearized equation of the form:
\begin{align}
    EIu''(s) + \kappa u(s) = 0 \label{eq:lin_elastica}
\end{align}
with the non-trivial solution:
\begin{align}
    u(s) = C \cos \frac{n \pi s}{l}, \text{ if and only if } \kappa = \frac{n^2 \pi^2 E I}{l^2}
    \label{eq:nontrivial_sol}
\end{align}
where $n$ is an integer. The lowest non-zero value $\kappa_{cr} = \pi^2 E I/l^2$ is the critical buckling load.

The solution to the linearized equation \eqref{eq:lin_elastica} gives some insight into the solution of \eqref{eq:elastica}, however, there are issues that cannot be addressed by the linearized analysis:
\begin{itemize}
    \item The necessary condition for bifurcation is the existence of a non-trivial solution near bifurcation point. However, this condition is not sufficient. Consider the nonlinear equation
    \begin{align} \label{eq:nonlinear_ex1}
        x^3 + \kappa x = 0
    \end{align}
    where $x$ is a real state variable and $\kappa$ is a real bifurcation parameter. The equation \eqref{eq:nonlinear_ex1} admits a trivial solution $x=0$ for all values of $\kappa$. The linearlized equation $\kappa x = 0$ has non-trivial solution $x=\const$ for $\kappa = 0$. The nonlinear equation, however, has no bifurcation solution branch at $\kappa = 0$.
    
    \item When a bifurcation branch exists, its qualitative behavior cannot be derived from the solution of the linearized equation. It is not possible to establish how many bifurcation branches there are and how these branches evolve as the bifurcation parameter varies. Consider, for illustration,
    \begin{align}
        x^3 - \kappa x = 0
    \end{align}
    which has a trivial solution $x=0$ for all $\kappa$. The linearized equation $\kappa x = 0$ has nontrivial solutions $x=\const$ near $\kappa = 0$. It is, however, not possible to determine the number of bifurcation branches nor their evolution post-bifurcation. The nonlinear equation on the other hand admits solutions $x=0$, $x = \sqrt{\kappa}$ and $x = -\sqrt{\kappa}$ when $\kappa > 0$ as post-bifurcation branches.

\end{itemize}

Singularity theory is developed to address these issues in a systematic fashion. 
It first employs the Lyapunov-Schmidt reduction to show that the solution to \eqref{eq:infinite_dim_equation}, which is defined in a function space in this case, is equivalent to that of an algebraic equation with one state variable. 
Next, by solving the recognition problem, the solution to the algebraic equation can be shown to be equivalent to that of a polynomial that exhibit the same bifurcation type in the neighborhood of an origin. 
Singularity theory can be expanded to bifurcation systems with symmetry. These equations presenting symmetry can be shown to be equivariant under certain group actions. Hence, the coupling of singularity theory and group theory provide tool for gaining insight into the persistence and/or the change of the symmetry that a bifurcation branch possesses near a bifurcation point \cite{chen2001singularity}.

The early concepts of singularity theory were first proposed by R. Thom, then developed rigorously by J. Mather \cite{mather1969stability, mather1970stability}, and later extended by V. I. Anorld \cite{arnold1976local, arnold1981singularity}. Subsequently, the development of singularity theory was systematized, then combined with group theory by M. Golubitsky, I. Stewart, and D. G. Schaeffer \cite{golubitsky2012singularitiesI,golubitsky2012singularitiesII}.

\subsection{Liapunov-Schmidt reduction}

Consider a smooth mapping $f\colon U \subset X \times K \to Y$ where $X$ and $Y$ are Banach spaces (i.e., complete normed vector spaces), $U$ is an open subset of $X$, and $K$ is an open subset of $\R^n$. The equation of interest is written as
\begin{align}
    f(u, \kappa) = 0
    \label{eq:Bif-1}
\end{align}
where $u$ determines the state of the system (e.g., deformation, temperature), $\kappa$ is a set of parameters (e.g., loads, geometry, material parameters), and $f$ is a nonlinear differential operator. $f$ is Fr\'echet-differentiable at $(u_0, \kappa_0) \in (U, K)$ with respect to $u \in U$ if there exists a bounded linear operator $D_u f(u_0, \kappa_0)\colon X \to Y$, referred to as the first-order Fr\'echet derivative, such that
\begin{align}
    \lim_{h \to 0} \frac{\| f(u_0 + h, \kappa_0) - f(u_0, \kappa_0) - h D_u f(u_0, \kappa_0) \|}{ \| h \|} = 0
    \label{eq:fr_deriv}
\end{align}
where $\| h \| = \|u - u_0 \|$ and $\| \cdot \|$ is a norm induced by an inner product on a Hilbert space which will be discussed below.
The mapping $f$ is assumed to be smooth in the sense that it admits a Fr\'echet derivative of any order. This definition of Fr\'echet derivative is akin to a Taylor expansion as $\| u - u_0 \| \to 0$ up to the first derivative. Higher order derivatives can be derived by extending the expansion to higher order terms. The implicit function theorem requires that $D_u f(u_0, \kappa_0) $ be invertible as a necessary condition for the existence of a bifurcation point at $(u_0, \kappa_0)$. 

We introduce Fredholm operators and Fredholm indexes. 
A bounded linear operator $L\colon X \to Y$ is called a \textit{Fredholm operator} if the kernel of $L$, defined as $\ker L \equiv \{u \in X\colon L(u) = 0 \}$, is a finite-dimensional subspace\footnote{A space (or subspace) is finite-dimensional if every element of that space (or subspace) can be represented as a finite linear combination of its basis. The dimension of the space (or subspace) is the number of elements in its basis.} of $X$ and the range of $L$, defined as $\range L \equiv \{y \in Y \colon L(u) = y  \text{ for some } u \in X\}$, is a closed subspace of $Y$ of finite-dimensional complement. If $L$ is a Fredholm operator, the \textit{Fredholm index} $i(L)$ is the integer
\begin{align}
    i(L) = \dim \ker L - \codim \range L
\end{align}
in which $\dim \ker L$ is the dimension of the kernel of $L$,  and $\codim \range L$, are the dimension of the complement the range of $L$ respectively. From the definition of Fredholm operators, $\dim \ker L$ and $\codim \range L$ are finite, hence the index $i(L)$ is finite. From this it follows that if $L\colon X \to Y$ is a Fredholm operator, then there exist closed subspaces $R$ and $Q$ of $X$ and $Y$ respectively, such that $X$ and $Y$ can be decomposed as follows:
\begin{subequations}
    \begin{equation}
        X = \ker L \oplus R 
        \label{eq:Decomp_a}
    \end{equation}
    \begin{equation}
        Y = Q \oplus \range L.  
        \label{eq:Decomp_b}
    \end{equation}
    \label{eq:Decomp}
\end{subequations}
in which we use the direct sum $\oplus$ of subspaces which allows to uniquely define an element of $X$ or $Y$ as a sum of elements of $\ker L$ and $R$, or $Q$ and $\range L$ respectively. In this paper, we will discuss only Fredholm operators with $i(L) = 0$ given that most problem in elasticity fall under this category. In this case, $\dim \ker L = \codim \range L = \dim Q$. For differential operators, it is typical to have Banach spaces that admit an inner product (i.e, Hilbert spaces). One such space is the Hilbert space $L^2(\Omega)$ where $\Omega$ is a bounded domain. The standard $L^2$ inner product is of the form  
\begin{align}
    \langle u, v \rangle = \int_{\Omega} uv
    \label{eq:inner_product}
\end{align}
which induces the norm $\| u \|^2 = \langle u, u \rangle$. For these spaces, the decompositions in \eqref{eq:Decomp} are orthogonal, i.e. $\langle u, v \rangle = 0$ for all $u \in \ker L$ and $v \in R$. Similarly, $\langle u, v \rangle = 0$ for all $u \in Q$ and $v \in \range L$.

We next describe the Lyapunov-Schimdt reduction of \eqref{eq:Bif-1}, which is assumed to admit a solution near $(0, \kappa_0) \in X \times K$. The differential operator $L \equiv D_u f(0, \kappa_0)$ is the Fr\'echet derivative; the Lyapunov-Schimdt reduction is applicable when the Fr\'echet derivative is a Fredholm operator at the bifurcation point. We assume a Fredholm operator with zero index throughout. We define an orthogonal projection $P\colon Y \to \range L $ and the complementary projection $(I - P) \colon Y \to Q$ from the split in \eqref{eq:Decomp_b}, where $I$ is the identity operator. This allows us to decompose \eqref{eq:Bif-1} into the following pair of equations:
\begin{subequations}
    \begin{equation}
        P f(u, \kappa) = 0 
    \end{equation}
    \begin{equation}
        (I - P) f(u, \kappa) = 0    
        \label{eq:system_eq_b}
    \end{equation}
    \label{eq:system_eq}
\end{subequations}
From the decomposition \eqref{eq:Decomp_a}, we write $u = v + w$ for some unique $v \in \ker L$ and $w \in M$. Then, we can define the map $G\colon \ker L \times R \times K \to \range L$ as 
\begin{align} 
    G(v, w, \kappa) \equiv P f(v + w, \kappa). 
    \label{eq:decompose} 
\end{align}
The Fr\'echet derivative $D_w G(v, w, \kappa)$ of $G$ with respect to $w$ is a linear map $R \to \range L$. In the neighborhood of $(0, \kappa_0)$, the bounded linear operator operator $D_w G(0, 0, \kappa)$ is a restriction of $L$ on $R$ and a bijection. 
For finite-dimensional spaces, the bijection of the differential operator implies it is invertible. In the case of Banach spaces, the additional condition that $\range L$ is closed, given that it is assumed to be a Fredholm operator, implies its invertibility. The projection of $u$ to $M$ and $f$ to $\range L$ factors out the invertible part of $f$. 

Before proceeding further, we introduce the implicit function theorem for Banach spaces due to its importance in determining bifurcation points. Consider $\phi\colon X \times K \to Y$, a $\mathcal{C}^1$ mapping (a mapping with continuous first derivative) between Banach spaces and let $D_u \phi(u, \kappa) \colon X \to Y$ be the Fr\'echet derivative of $\phi$ with respect to $u$ as defined in \eqref{eq:fr_deriv}. The \textit{implicit function theorem} states that for a $\mathcal{C}^1$ mapping $\phi$ near a fixed point $(u_0, \kappa_0)$ defined above, and supposing that $D_u \phi(u_0, \kappa_0)$ has a bounded inverse, i.e., $D_u \phi(u_0, \kappa_0) \neq 0$, then the equation $\phi(u_0, \kappa_0) = 0 $ can be solved locally, in the neighborhood $\mathcal{N}(\kappa_0)$ of $\kappa_0$ in $\ker L$, for $u_0=W(\kappa_0)$, where $W\colon \mathcal{N}(\kappa_0) \to X$ is a $\mathcal{C}^1$ function. In essence, the Lyapunov-Schmidt reduction makes it possible to apply the implicit function theorem to equations where it is not readily applicable as is the case near bifurcation points where the Fr\'echet derivative is not invertible. 

Returning to the map $G$ in \eqref{eq:decompose} and its Fr\'echet derivative $D_w G(v, w, \kappa)$, which we have argued is invertible, we apply the implicit function theorem to solve for a unique $w$ in the neighborhood $\mathcal{N}(0, \kappa_0)$ of the form $w = W(v, \kappa)$ with $W\colon \ker L\times K \to R$  which satisfies:
\begin{subequations}
    \begin{equation}
        P f(v + W(v, \kappa), \kappa) = 0,
    \label{eq:projection}   
    \end{equation}
    \begin{equation}
        W(0, \kappa_0) = 0.
        \label{eq:condition_w}
    \end{equation}
\end{subequations}
    
This solution is substituted in \eqref{eq:system_eq_b} to give:
\begin{align}
    g(v, \kappa) = (I -P) f(v + W(v, \kappa), \kappa).
\end{align}
From the solution $W(0, \kappa_0) = 0$, it follows that $g(0, \kappa_0) = 0$.
The essential result of the Lyapunov-Schmidt reduction is that if the differential operator in the linearization of \eqref{eq:Bif-1} is a Fredholm operator of index zero, then solutions of \eqref{eq:Bif-1} are in one-to-one correspondence with 
\begin{align}
    g(v, \kappa) = 0
    \label{eq:reduced_eq}
\end{align}
in the neighborhood of $(0, \kappa_0)$. \eqref{eq:reduced_eq} is referred to as reduced bifurcation equation.

In the next step, we will choose a basis $e_1, \ldots, e_n$ for $\ker L$ and $e*_1, \ldots, e*_n$ for $Q$. This choice is possible because both subspaces are Banach spaces and also because we assumed $L$ to be a Fredholm operator of index zero, which leads to finite-dimensional subspaces $\ker L$ and $Q$ with equal dimension $n$. An element in $\ker L$ can be written as $v = v_i e_i$, using the Einstein summation convention. 
Then, the inner product of \eqref{eq:reduced_eq} with each $e*_i$ forms a system of equations:
\begin{equation}
\begin{split}
    g_i(v, \kappa) 
        &\equiv \langle e*_i, (I -P) f(v + W(v, \kappa), \kappa) \rangle \\
        &= \langle e*_i, f(v + W(v, \kappa), \kappa) \rangle \\ 
        &= \langle e*_i, f(v_j e_j + W(v_j e_j, \kappa), \kappa) \rangle
\label{eq:w_basis}
\end{split}
\end{equation}
with the second equation arising from the fact that the projection $P$ maps $f$ to $\range L$ which is orthogonal to $Q$, and $i, j =1, \ldots, n$. \eqref{eq:w_basis} is equivalent to \eqref{eq:reduced_eq}, and hence the bifurcation equation can be rewritten as:
\begin{align}
    g_i(v, \kappa) = 0
\end{align}
and from \eqref{eq:Bif-1}, \eqref{eq:condition_w} and \eqref{eq:w_basis}, satisfying 
\begin{align}
    g_i(0, \kappa_0) = 0.
\end{align}
To solve \eqref{eq:w_basis}, it is necessary to determine low-order terms of its expansion. We will compute the derivative with respect to $v_j$ and $\kappa$. The derivative of $W$ will be determined by an implicit differentiation of \eqref{eq:projection} and the chain rule, and these will be used to compute the derivatives of \eqref{eq:w_basis} near $(v, \kappa) = (0, \kappa_0)$. 

We substitute $v = v_i e_i$ in \eqref{eq:projection} and differentiate with respect to $v_i$ to get
\begin{align}
    P L(e_i + W_{v_i}) = 0, \quad \text{ where } W_{v_i} := \frac{\partial W}{\partial v_i}.
    \label{eq:deriv_wi}
\end{align}
For conciseness, subscripts will be used to denote partial derivatives below. From the linearity of $P$, \eqref{eq:deriv_wi} implies that $L(e_i + W_i) = 0$, thus $e_i + W_{v_i} \in \ker L$. However, $W_v \in R$ by definition, hence
\begin{align}
    W_{v_i}(0, \kappa_0) = 0.
    \label{eq:w_deriv_wi0}
\end{align}
Next, we differentiate \eqref{eq:projection} with respect to $\kappa$ and use the linearity of $P$ to get
\begin{align}
    P LW_{\kappa} + Pf_{\kappa} = 0 \quad \implies \quad  W_{\kappa} = -L^{-1} Pf_{\kappa} 
    \label{eq:w_deriv_k}
\end{align}
where $L^{-1}$ is the inverse of the differential operator restricted to $R$ and $f_{\kappa}$ is the Fr\'echet derivative of $f$ with respect to $\kappa$ evaluated at $(u, \kappa) = (0, \kappa_0)$. Similarly, differentiating \eqref{eq:projection} with respect to $v_i$ and $v_j$ gives:
\begin{align}
    W_{v_i v_j} = -L^{-1} Pf_{uu}e_i e_j.  
\end{align}
Using these results, we now compute the derivatives of \eqref{eq:w_basis}, taking into account the fact that $e*_i$ is orthogonal to $\range L$. All derivatives will be taken at $(v, \kappa) = (0, \kappa_0)$. 
Differentiating \eqref{eq:w_basis} with respect to $v_j$, we get
\begin{align}
    {g_i}_{v_j} =  \langle e*_i, L(e_j + W_{v_j}) = 0 \rangle . 
\end{align}
Similar calculations employing the results from \eqref{eq:w_deriv_wi0} and \eqref{eq:w_deriv_k} give:
\begin{align}
    {g_i}_{\kappa} & =  \langle e*_i, LW_{\kappa} + f_{\kappa}\rangle = \langle e*_i, f_{\kappa}\rangle,  
    \\
    {g_i}_{v_j v_k} & =  \langle e*_i, f_{uu} (e_j + W_{v_j}) (e_k + W_{v_k}) + LW_{v_j v_k}\rangle = \langle e*_i, f_{uu} e_j e_k\rangle,  
    \\
    {g_i}_{v_j v_k} & =  \langle e*_i, f_{uu} (e_j + W_{v_j}) (e_k + W_{v_k}) + LW_{v_j v_k}\rangle = \langle e*_i, f_{uu} e_j e_k\rangle,  
    \\
    \begin{split}
        {g_i}_{v_j \kappa} 
            &=  \langle e*_i, f_{uu} (e_j + W_{v_j}) W_{\kappa} + f_{u \kappa}(e_j + W_{v_j}) + L W_{v_j \kappa}\rangle \\
            &= \langle e*_i, -f_{uu} e_j (L^{-1} Pf_{\kappa}) + f_{u \kappa} e_j\rangle
    \end{split}
    \\
    \begin{split}
        {g_i}_{v_j v_k v_l} 
            & =   \langle e*_i,  f_{uuu} (e_j + W_{v_j}) (e_k + W_{v_k})(e_l + W_{v_l})  \\ 
            & \quad + f_{uu} [(e_k + W_{v_k}) W_{v_j v_l} + (e_j + W_{v_j}) W_{v_k v_l} + (e_l + W_{v_l}) W_{v_j v_k}] + LW_{v_j v_k v_l}\rangle \\
            & = \langle e*_i, f_{uuu} e_j e_k e_l + f_{uu} (e_k W_{v_j v_l} + e_j W_{v_k v_l} + e_l W_{v_j v_k})\rangle \\ 
            & = \langle e*_i, (f_{uuu} - 3 f_{uu} L^{-1} Pf_{uu}) e_j e_k e_l \rangle. 
    \end{split}
\end{align}

\subsubsection{Example: Bifurcation in the Elastica Problem}

For illustration, we return to the elastica problem \eqref{eq:elastica} with boundary conditions \eqref{eq:elastica_bc}. We reformulate the problem to define the various Banach spaces and the bifurcation problem \eqref{eq:Bif-1}. Let 
\begin{align*}
    U \subset X \equiv \{ u \in \mathcal{C}^2([0, l]; \R) : u'(0) = u'(l) = 0\}, \quad Y \equiv \mathcal{C}^0([0, l]; \R), \quad K = \R,
\end{align*}
where $\mathcal{C}^n([0, l]; \R)$ is the space of real-valued, $n$-continuously differentiable functions (when $n=0$, the function is not continuously differentiable), and let
\begin{align}
    f(u(s), \kappa) = EIu''(s) + \kappa \sin u(s).
\end{align}
The Fr\'echet derivative $L$ of $f$ with respect to $u$ at $(u, \kappa) = (0, \kappa_0)$ is given by
\begin{align}
    L u(s) = EIu''(s) + \kappa_0 u(s).
\end{align}
The solution at equilibrium is found by solving the boundary value problem $Lu = 0$ for $u \in X$ to obtain: 
\begin{equation}
    \dim \ker L = 
    \begin{cases}
        1, & \text{ if } \kappa_0 = n^2 \pi^2 EI/l^2 \\
        0, & \text{ otherwise.} 
    \end{cases}
\end{equation}
We will be concerned with the case $\kappa_0 = n^2 \pi^2 EI/l^2$, where $\ker L$ is given by:
\begin{align}
    \ker L = \{ u \in X\colon u(s)=C\cos{\frac{n \pi s}{l}}, C \in \R\} .
    \label{eq:kerL}
\end{align}
The orthogonal complement $R$ of $\ker L$ in $X$ is then given by 
\begin{align}
    R = \left\{ w \in X\colon \int_0^l w(s) \cos \frac{n \pi s}{l} \dm s = 0 \right\} \nonumber.
\end{align}
where we employed the inner product \eqref{eq:inner_product} equal to zero to define the orthogonal complement. Similarly, the orthogonal complement $Q$ to $\range L$ in $Y$ contains elements $y(s)$ that, for every $u \in X$, can be obtained from:
\begin{equation*}
\begin{split}
    \langle y, Lu \rangle 
        &= \int_0^l y(s)[EI u''(s) + \kappa_0 u(s)] \dm s  \\
        &= [EI y(s) u'(s)]_0^l + \int_0^l [EI y'(s) u'(s) + \kappa_0 y(s) u(s)] \dm s  \\ 
        &= [EI y'(s) u(s)]_0^l + \int_0^l [EI y''(s) + \kappa_0 y(s)] u(s) \dm s  \\
        &= \langle Ly, u \rangle = 0    
\end{split}
\end{equation*}

where we integrated by parts twice and used the boundary conditions \eqref{eq:elastica_bc}. The orthogonality condition above implies that $y(s) \in Q$ must satisfy:
\begin{align}
    &L y  = EI y''(s) + \kappa_0 y(s) = 0 \quad \text{on} ~(0,l) \\
    &y'(0) = y'(l) = 0.
\end{align}

The equation above implies that the elements $y$ belong to $\ker L$, i.e $Q =  \ker L$. This in turn implies that the orthogonal complement $R$ of $\ker L$ is $\range L$. 
The subspaces $\ker L$ and $Q$ are spanned by the bases 
 \begin{align}
     e = e^* = \sqrt{\frac{l}{2}} \cos{ \frac{n \pi s}{l}}
 \end{align}
derived from \eqref{eq:kerL} with a normalization coefficient.

The projection $Py(s)$ of the space $Y$ onto the subspace $\range L$ is obtained by subtracting the elements of $\ker L$ from elements of $Y$. This is possible because we have shown that the subspaces $\range L$ and $\ker L$ are orthogonal complements of $Y$. Hence,
\begin{align*}
     Py(s) = y(s) - \left[\frac{2}{l} \int_0^l y(t) \cos{\frac{n \pi t}{l}} \dm t \right] \cos{\frac{n \pi s}{l}}.
\end{align*}

\subsection{Recognition problem}

The singularity theory approach to bifurcation problems focuses on two issues.
First, the importance of higher order derivatives in the Taylor expansion of \eqref{eq:Bif-1} in determining the qualitative behavior of the solution. In other words, to what extent is the qualitative behavior determined by low-order derivatives of \eqref{eq:Bif-1}? The singularity theory term for this problem is \textit{finite determinacy}. 
Second, finding a polynomial equation as simple as possible whose solution is in one-to-one correspondence with the given equation near a bifurcation point. 
This is termed as the \textit{recognition problem} and the polynomial is referred to as the \textit{normal form} which can be determined by a finite number of derivatives of the given equation. 
The normal form will have the same qualitative behavior as the reduced bifurcation equation near the bifurcation. 

Here, we focus on solving recognition problem for the reduced bifurcation problem in \eqref{eq:reduced_eq} of the form $g(x, \kappa)$.
We choose a 1-d space ($\R$) for both the state variable and bifurcation parameter, and choose the bifurcation point to be at the origin $(0,0)$ for convenience without loss of generality.

We define two smooth mappings $g, h \colon \mathcal{N} \times \R \to \R$ as \textit{strongly equivalent} if there exist functions $X(x, \kappa)$ and $S(x, \kappa)$ such that the relation
\begin{subequations}
    \begin{equation}
        g(x, \kappa) = S(x, \kappa) h(X(x, \kappa), \kappa),
        \label{eq:strong_equiv}
    \end{equation}
    holds near the origin and that the conditions
    \begin{equation}
        X(0, 0) = 0, \quad X_x(x, \kappa) > 0, \quad S(x, \kappa) > 0
        \label{eq:strong_equiv_cond}
    \end{equation}
\end{subequations}
are satisfied.
The most important consequence of equivalence is that the number of solutions of \eqref{eq:reduced_eq} is preserved when this equation is replaced by
\begin{align}
    h(X, \kappa) = 0.
\end{align}
To prove this, suppose that for a given $\kappa$, the reduced bifurcation equation $g(x, \kappa)$  has exactly $n$ solutions of the form $x_1 < x_2 < \ldots < x_n$ such that 
\begin{align}
    g(x_i, \kappa) = 0 ~ \text{for all}~i=1,\ldots,n.
\end{align}
From \eqref{eq:strong_equiv} and \eqref{eq:strong_equiv_cond}$_3$, we have: 
\begin{align}
    h(X, \kappa) = 0 ~ \text{if and only if} ~ X = X(x_i, \kappa), i =1,\ldots,n.
\end{align}
And it is obvious from \eqref{eq:strong_equiv_cond}$_2$ that $X(x_1, \kappa) < X(x_2, \kappa) < \ldots < X(x_n, \kappa)$.

The main idea behind the recognition problem is to  explicitly characterize the smooth mappings which are strongly equivalent to $g$ near the origin. 
Much of singularity theory is devoted to finding the simplest normal form for a certain function for which a number of derivatives are given or computed at a certain bifurcation point. It is worth mentioning that for more complicated problems, with various degrees of sophistication, it may not be practical to derive the normal form from the derivatives. \cite{chen2001singularity} gives examples of some bifurcation problems whose normal form can be derived from the derivatives. In the next section, we will demonstrate, by using elementary calculus, how a normal form can be constructed and some conditions of equivalence between two functions are established through the examination of their tangent spaces. 
 
 \subsubsection{Recognition problem for pitchfork bifurcation}
 The main focus for this section will be to determine the conditions under which the reduced function $g(x, \kappa)$ is strongly equivalent to $h(x, \kappa)$. 

Let $\mathcal{E}_{x, \kappa}$ denote a space of smooth functions $g\colon \R^2 \to \R$ on some neighborhood of the origin.

\begin{definition}[Restricted tangent space]

Let $g \in \mathcal{E}_{x, \kappa}$. A function $f(x, \kappa)$ belongs to the restricted tangent space of $g(x, \kappa)$, denoted as RT($g$) if and only if there exist smooth functions $a(x, \kappa), b(x, \kappa), c(x, \kappa) \in \mathcal{E}_{x, \kappa}$ such that
\begin{align}
    f(x, \kappa) = a g + (x b + \kappa c) g_x,
    \label{eq:tangent_space}
\end{align}
where $g_x$ denotes the partial derivative of $g$ with respect to $x$.
The notion of restricted tangent space is linked to the notion of strong equivalence introduced in the previous section by the fact that the function $f$ is said to be in RT$(g)$ if $g + \eps f$ is strongly equivalent to $g$ for all small $\eps$. To show this, suppose that for some perturbation $f$, the strong equivalence is satisfied. Then, for some small $\eps$, there exist $S(x, \kappa, \eps)$ and $X(x, \kappa, \eps)$ such that 
\begin{align}
    g(x, \kappa) + \eps f(x, \kappa) = S(x, \kappa, \eps) g (X(x, \kappa, \eps), \kappa),
    \label{eq:equivalence_f}
\end{align}
where $X(0, 0, \eps) \equiv 0$ at the origin. Suppose further that both $S$ and $X$ are smooth functions of $x, \kappa$ and $\eps$, and that at $\eps=0$, the identity transformation for $g$ is such that 
\begin{align}
    S(x, \kappa, 0) \equiv 1, \quad X(x, \kappa, 0) \equiv x.
\end{align}
It follows then, by differentiating \eqref{eq:equivalence_f} with respect to $\eps$:
\begin{align}
    f(x, \kappa) = S_{\eps} (x, \kappa, 0) g(x, \kappa) + g_x (x, \kappa) X_{\eps} (x, \kappa, 0),
\end{align}
where we identify the terms $a = S_{\eps} (x, \kappa, 0)$ and $xb + \kappa c = X_{\eps} (x, \kappa, 0)$ in \eqref{eq:tangent_space}.
The restricted tangent space defines a necessary condition for strong equivalence. 
\end{definition}

To illustrate the idea, we use a simple example of a pitchfork bifurcation $g = \kappa x - x^3$, and determine the necessary conditions on the strongly equivalent function $f(x, \kappa)$. To achieve this, we use the definition in \eqref{eq:tangent_space}, and group terms to obtain
\begin{align}
    f(x, \kappa) = a ( \kappa x - x^3) + (x b + \kappa c)( \kappa - 3 x^2) = (a + b -3xc) \kappa x - (a + 3b) x^3  + c \kappa^2,
    \label{eq:equiv_pitchfork}
\end{align}
where $a, b$ and $c \in \mathcal{E}_{x, \kappa}$. We then write \eqref{eq:equiv_pitchfork} as: 
\begin{align}
    f(x, \kappa) = \alpha(x, \kappa) x^3 + \beta(x, \kappa) \kappa x + \gamma(x, \kappa) \kappa^2,
    \label{eq:equiv_pitchfork2}
\end{align}
where:
\begin{align*}
    -a - 3b &= \alpha, \\
    a + b - 3xc &= \beta, \\
    c &= \gamma.
\end{align*}
Finally, we apply a Taylor expansion of $f$ near the origin $(x, \kappa) = (0, 0)$:
\begin{align}
   f(x, \kappa) = f(0,0) + f_x(0,0) x + f_{\kappa} (0,0) \kappa + f_{xx}(0,0) x^2 + f_{x \kappa} (0,0) x \kappa + f_{\kappa \kappa} (0,0) \kappa \kappa + f_{xxx} (0,0) x^3
   \label{eq:taylor_exp}
\end{align}
Comparing \eqref{eq:taylor_exp} and \eqref{eq:equiv_pitchfork2} gives:
\begin{align}
    f(0,0) = f_x(0,0) = f_{\kappa} (0,0) = f_{xx}(0,0) = 0,
    \label{eq:equiv_condition}
\end{align}
and
\begin{align*}
    \alpha = f_{xxx} (0,0), ~ \beta = f_{x \kappa} (0,0), ~\text{and}~ \gamma =  f_{\kappa \kappa} (0,0).
\end{align*}
The conditions in \eqref{eq:equiv_condition} are referred to as the \textit{defining conditions} of the normal form. 
Additional conditions, in the form of inequalities, are needed to establish that a normal form is strongly equivalent to a reduced equation. 
These conditions are will be referred to as \textit{nondegeneracy conditions}. 

 \eqref{eq:equiv_pitchfork2} can be simplified further to $ h(x, \kappa) = \alpha(x, \kappa) x^3 + \beta(x, \kappa) x \kappa$ by redefining $f(x, \kappa) = h(x, \kappa) + \gamma(x, \kappa) \kappa^2$, then employing Theorem 2.2 in 
 \cite{golubitsky2012singularitiesI} which states that given $f, p \in \mathcal{E}_{x, \kappa}$; if $\text{RT}(f + tp) = ~\text{RT}(f)$ for all $t \in [0,1]$, then $f + tp$ is strongly equivalent to $f$ for all $t \in [0,1]$. It suffices to show that $\text{RT}(f) = ~\text{RT}(h)$ to established the strong equivalence between the two functions.
We follow \cite{chen2001singularity} by showing that  $\text{RT}(h) \subset ~\text{RT}(f)$, using the definition \eqref{eq:tangent_space}, such that,
\begin{align}
    a f + (x b + \kappa c) f_x 
        &= a (h + \gamma \kappa^2) + (x b + \kappa c) (h_x + \gamma_x \kappa^2) \\
        &= \overline{a} h + (\overline{b} x + \overline{c} \kappa) h_x
\end{align}
where $\overline{a}, \overline{b}$ and $\overline{c}$ are smooth functions given by the expressions
\begin{align*}
    \overline{a} &:= a - \frac{[a \gamma + (b x + c \kappa) \gamma_x](3 \alpha + \alpha_x x)^2 x}{[2 \alpha \beta + (\beta \alpha_x - \alpha \beta_x )x](\beta + \beta_x x)}, \\
    \overline{b} &:= b + \frac{\alpha [a \gamma + (b x + c \kappa) \gamma_x](3 \alpha + \alpha_x x) x}{[2 \alpha \beta + (\beta \alpha_x - \alpha \beta_x )x](\beta + \beta_x x)},\\
    \overline{c} &:= c + \frac{a \gamma + (b x + c \kappa) \gamma_x}{ \beta + \beta_x x}.
    \label{eq:equiv_transf}
\end{align*}

The degeneracy conditions can be derived by comparing the sign of coefficients of $\kappa x$ and $x^3$ in the normal form $g = \kappa x - x^3$ and \eqref{eq:taylor_exp} which leads to the following:
\begin{align}
    \alpha = f_{xxx} < 0 , ~\beta = f_{\kappa x} > 0.
\end{align}

In the example, we used elementary calculus to derive the defining and degeneracy conditions for the pitchfork bifurcation. Chief among these is the Taylor expansion of the smooth function $g$, which consist of a linear expansion of the partial derivatives with respect to both the state variable and the bifurcation parameter in the neighborhood of the origin. These partial derivatives of $g$ can be grouped into three classes, namely, low-, intermediate- and high-order terms. These play an important role in solving the recognition problem. It is worth mentioning here that we have only considered a simple bifurcation problem. For more general bifurcation problems, sophisticated techniques from group theory and algebra might be needed, as described in \cite{golubitsky2012singularitiesI} and \cite{golubitsky2012singularitiesII} to solve the recognition problem. We will restrict our focus to the simple cases while explaining, briefly and without proof, some concepts necessary for understanding the procedure. 

 The low-order terms in the  Taylor expansion are the monomials of the form $x^p \kappa^q$ such that the corresponding partial derivatives $\partial^{p+q} g / \partial x^p \partial \kappa^q = 0$ in the defining conditions. In the example above, the low order terms will be $1, x, \kappa$ and $x^2$ associated with the partial derivatives in \eqref{eq:equiv_condition}. These terms are excluded from the Taylor expansion and the subsequent equivalence transformation in  \eqref{eq:equiv_transf}. 

 The higher order terms of the expansion are those that can be transformed out of the expression through strong equivalence as was done in \eqref{eq:equiv_transf}. For our example, these are $\kappa^2, \kappa x^2, \kappa^2 x, \kappa^3, \ldots$. These can be identified through an analysis of the normal form and its perturbations employing the theorem stated above. These terms can be incorporated with other terms in the Taylor expansion after an appropriate redefinition of the coefficients. It is also worth mentioning that the higher-order terms do not correspond, in general, to monomials $x^p \kappa^q$ of the Taylor expansion whose associated derivatives $\partial^{p+q} g / \partial x^p \partial \kappa^q$ are not in neither the defining nor the degeneracy conditions. Some higher order monomials will be absorbed into other monomials through the equivalence transformation; thus they will not appear in the normal form even though their associated partial derivatives appear in either the defining or nondegeneracy conditions.

 The monomial terms that are neither low-order nor higher-order terms are those that remain in the normal form. These are referred to as intermediate-order terms. After the given function is reduced to finite intermediate-order terms, it only remains to reduce their coefficients to constants to get the final expression of the normal form. 
 Chapter 9 of \cite{chen2001singularity} describes several solutions to the recognition problems for various bifurcation problems, as well their defining and nondegeneracy conditions. 

In the section that follows, a pitchfork bifurcation problem with $Z_2$ symmetry is analyzed, in the context of an homogeneous elastic body in the presence of an electric field.

\section{Bifurcation Analysis} \label{Analysis_Results}

The instability in the T-K problem was previously studied by Chen \cite{chen1991bifurcation,MACSITHIGH1992} for a homogeneous elastic body under dead-load tractions with $Z_2$ symmetry. 
This symmetry is present in our problem with only out-of-plane electric fields\footnote{
    As shown in \cite{katusele2025exploiting}, in-plane fields break this symmetry.
} and with uniform loading $\hT_1 = \hT_2 = \hT$ in \eqref{str1} and \eqref{str2}.
This is  demonstrated by the invariance of $\hpsi(\la_1, \la_2, \hE_3)$ in \eqref{comp_engy_lambda} under the permutation of $\la_1$ and $\la_2$, in the principal plane of $F$, i.e.,
\begin{align}
    \hpsi (\la_1, \la_2, \hE_3) = \psi (\la_2, \la_1, \hE_3) 
    \label{eq:symmetry}
\end{align}

We denote symmetric solutions to \eqref{str1} and \eqref{str2} by $\la = \la_1 = \la_2$ and use the change of variables:
\begin{equation}
x = \frac{\la_1 + \la_2}{2} - \la, \quad y = \frac{\la_1 - \la_2}{2}, \quad \text{and} \quad \tau = \hT - \that_0.
\end{equation} 
In the case of symmetric stretching, the solution will be obtained for values of $x = y = \tau = 0$. 
We highlight that $y$ measures the departure from symmetric deformation. 
We rewrite the equilibrium equations by summing and subtracting, respectively, \eqref{str1} and \eqref{str2} to get:
\begin{align}
    \hT_1 (\lambda + x + y, \lambda + x - y, \hE_3) + \hT_2 (\lambda + x + y, \lambda + x - y, \hE_3) - 2(\hT + \tau) = 0 \label{eq:transformation1}\\
    \hT_1 (\lambda + x + y, \lambda + x - y, \hE_3) - \hT_2 (\lambda + x + y, \lambda + x - y, \hE_3) = 0 \label{eq:transformation2}
\end{align}

It then follows from \eqref{eq:symmetry} that the left sides of \eqref{eq:transformation1} and \eqref{eq:transformation2} are, respectively, even and odd in $y$.
From a theorem by \cite{whitney1943differentiable}, it follows that there exist smooth functions $p(x,y^2,\tau)$ and $y q(x,y^2)$, such that:
\begin{align}
    p(x, y^2, \tau) = \hT_1 (\lambda + x + y, \lambda + x - y, \hE_3) + \hT_2 (\lambda + x + y, \lambda + x - y, \hE_3) - 2(T + \tau) \\
    yq(x, y^2) = \hT_1 (\lambda + x + y, \lambda + x - y, \hE_3) - \hT_2 (\lambda + x + y, \lambda + x - y, \hE_3).
\end{align}
With this reformulation, \eqref{eq:transformation1} and \eqref{eq:transformation2} can be written as
\begin{align}
    g(x, y, \tau) := (p(x, y^2, \tau), y q(x, y^2))  = 0,
\end{align}
where $g$ is defined in the neighborhood $\mathcal{N}$ of $\R^2 \times \R$ similar to that introduced in the previous section. 

We then compute the first order partial derivatives at $x = y = \tau = 0$:
\begin{align} 
\label{def_cond}
    p &= 0  & q= 2 + \frac{2}{\la^6} + \frac{6 \ga}{\la^4} - 2\la^2(\ga - \hE_3^2) 
    \\
    p_x & := \frac{\partial p}{\partial x} = 2 + \frac{10}{\la^6} + \frac{6\ga}{\la^4} + 6 \la^2 \left(\ga - \hE_3^2\right) 
           & q_x :=\frac{\partial q}{\partial x} = -\frac{12}{\la^7} - \frac{24 \ga}{\la^5} -4 \la \left(\ga - \hE_3^2 \right)
    \\
    p_{y^2} & :=\frac{\partial p}{\partial\left( y^2\right)} = -\frac{12 \ga}{\la ^5}-2 \la  \left(\ga -\hE_3^2\right)-\frac{6}{\la ^7} 
          & q_{y^2} :=\frac{\partial q}{\partial \left(y^2\right)} = \frac{6}{\la^8} + \frac{20 \ga}{\la^6} + 2 \left(\ga - \hE_3^2\right) 
    \\
    p_{\tau} & := \frac{\partial p}{\partial \tau} = -2 
\end{align}
The derivative $p_x$ is positive when $\hE_3 < \sqrt{\ga}$ and $q$ is monotone decreasing in the same interval. The solution of the recognition problem associated with the $Z_2$ equivariant function gives that $p=q=0$ is the defining condition for bifurcation. The stretch corresponding to the bifurcation point is the root of $q=0$ and has the form 
\begin{align}
    \lab^2 = -\frac{1}{4 \left(\hE_3^2-\ga \right)} + \half \sqrt{\frac{1}{4 \left(\hE_3^2-\ga \right)^2} + A + B} + \half \sqrt{\frac{1}{2 \left(\hE_3^2-\ga \right)^2} - A - B + \frac{1 + 24 \ga \left(\hE_3^2-\ga \right)^2 }{\left(\hE_3^2-\ga \right) \sqrt{1 + 4(A + B) \left(\hE_3^2-\ga \right)^2 }}}
\end{align}
where
\begin{align*}
 A &= \frac{\sqrt[3]{2} \left(4 \hE_3^2-7 \ga \right)}{\left(\ga -\hE_3^2\right) \left(243 \ga ^2 \left(\ga -\hE_3^2\right)+\sqrt{\left(243 \ga ^2 \left(\ga -\hE_3^2\right)-27\right)^2-4 \left(12 \hE_3^2-21 \ga \right)^3}-27 \right)^{1/3}}   
 \\ 
 B &= \frac{\left(243 \ga ^2 \left(\ga -\hE_3^2\right)+\sqrt{\left(243 \ga ^2 \left(\ga -\hE_3^2\right)-27\right)^2-4 \left(12 \hE_3^2-21 \ga \right)^3}-27\right)^{1/3}}{3 \sqrt[3]{2} \left(\ga -\hE_3^2\right)}.
\end{align*}
The critical bifurcation load $\thatb$ computed by substituting $\la_1 = \la_2 = \lab$ in either \eqref{str1} or \eqref{str2} is of the form
\begin{align}
    \thatb = \lab + \left( \ga -\hE_3^2 \right) \lab ^3  - \left( \ga  \lab ^{2} + 1 \right) \lab ^{-5}
    \label{critical_load}
\end{align}

\begin{figure}[htb!]
    \centering
    \subfloat[]{\includegraphics[width=0.49\textwidth]{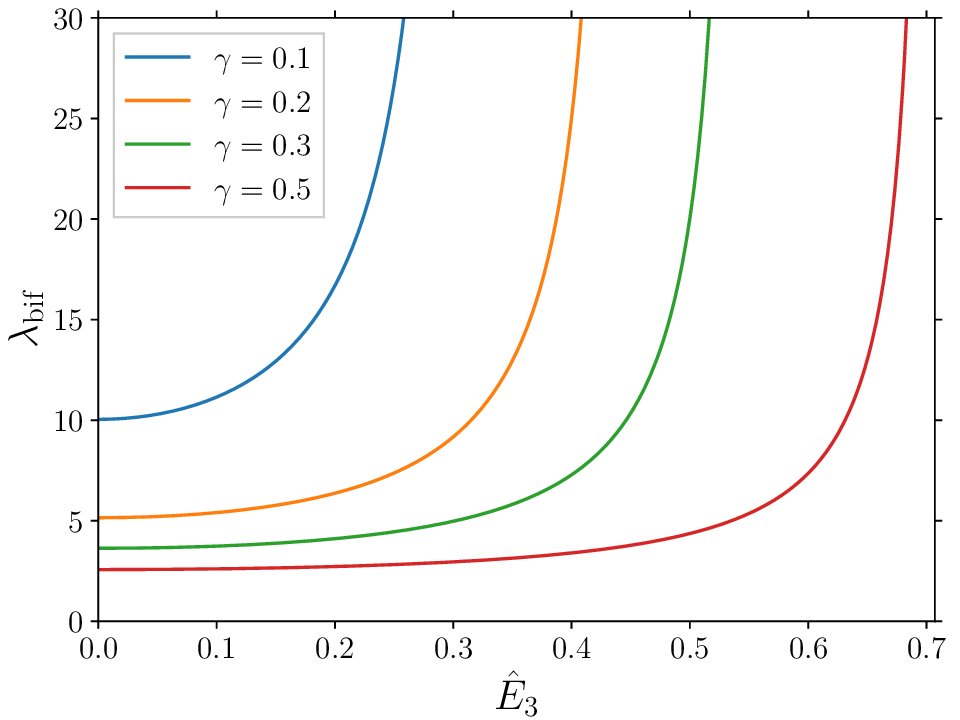}\label{fig:bif_pt}}
    \subfloat[]{\includegraphics[width=0.49\textwidth]{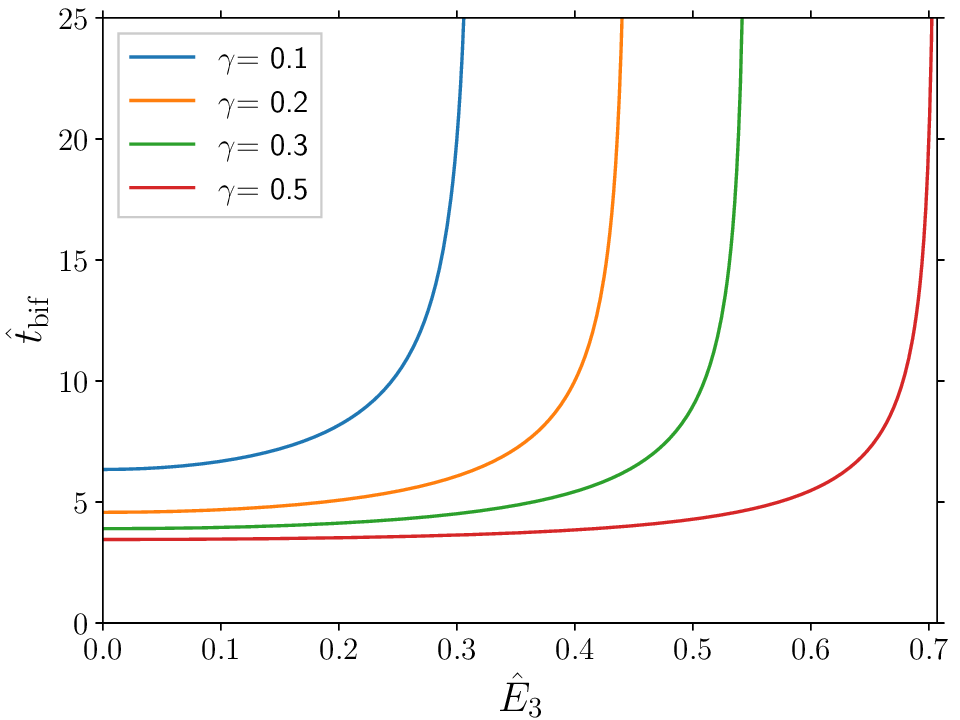} \label{fig:bif_pt2}}
    \caption{(a) Bifurcation stretch $\lab$  as a function of the through-thickness electric field $\hE_3$. (b) Critical load $\thatb$ at bifurcation as a function of $\hE_3$. The bifurcation stretch and critical load go to infinity as $\hE_3 \to \sqrt{\ga}$ from below. }
\end{figure}

A plot of $\lab$ against $\hE_3$ in Figure \ref{fig:bif_pt} shows a nonlinear increase in the critical bifurcation stretch as the electric field increases. 
This means the load required to produce an asymmetric deformation increases with increase in electric field as shown in Figure \ref{fig:bif_pt2}. 
The limiting value beyond which no bifurcation will occur (the stretch at bifurcation goes to infinity) is when $\hE_3 \ge \sqrt{\ga}$. The effect of the material parameter $\ga$ on stretch and critical load is plotted in Figures \ref{fig:bif_pt} and \ref{fig:bif_pt2} respectively. It is observed that increasing $\ga$ increases the limiting value of $\hE_3$ needed to suppress bifurcation. For an application where the T-K instability needs to be suppressed, a material should be selected to minimize $\ga$ for better performance.

\subsection{Linear Approximation of the Post-Bifurcation Behavior}

We now approximate the bifurcation curve by using a perturbation approach, near the origin (with no applied traction) for the pre-bifurcation branch and near the bifurcation point for the post-bifurcation branch. 
In the pre-bifurcation branch, the curve is matched with the solution for the unloaded state as well as the deformed state at the bifurcation point. 

\begin{figure*}[thb!]
    \centering
    \includegraphics[width=0.6\textwidth]{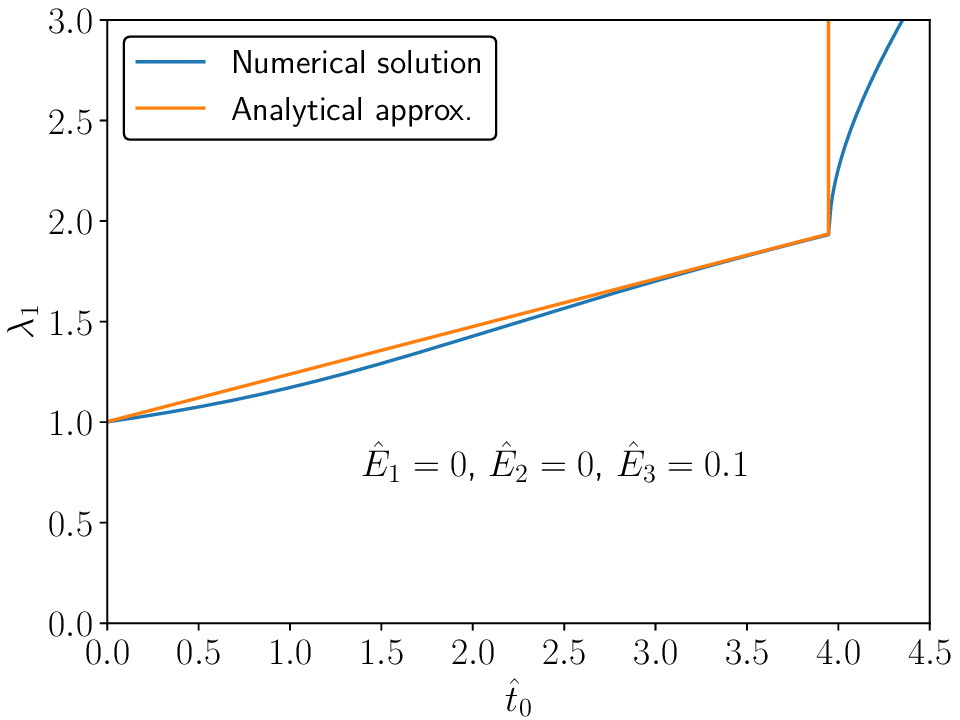} 
    \caption{Comparison between the numerical solution and the perturbation approximation for the bifurcation diagram.}
    \label{fig:bifurcation-approx}
\end{figure*}

We consider the perturbation $\la_1 = \la_2 = 1 + \epsilon$ where $\epsilon$ is a small parameter. 
Substituting this in \eqref{str1}, and eliminating terms that are higher-order in $\epsilon$, we get:   
\begin{equation} \label{eq_eps}
    - \that_0 - \hE_3^2 + 3\epsilon (2 + 2 \ga - \hE_3^2)= 0
\end{equation} 
We solve for $\epsilon$ in \eqref{eq_eps} and get the stretch:
\begin{equation} \label{eq_eps2}
    \la_1 = \la_2 = 1 + F\frac{\that_0 + \hE_3^2}{3(2 + 2 \ga - \hE_3^2)}
\end{equation} 
which matches the slope of the stretch-load curve near $\la_1 = \la_2 = 1$ and $\that_0 = 0$ when the factor $F = 1$. 
To further match the bifurcation point with \eqref{eq_eps2}, $F$ is computed to be:
\begin{equation} 
     F = \left(\lab -1\right) \frac{3 \left(2 + 2 \ga -\hE_3^2\right)}{\that_{\text{bif}}+\hE_3^2}
\end{equation}
where $\thatb$ is the critical bifurcation load.
We see in Figure \ref{fig:bifurcation-approx} that this approximation matches well with the numerical solution, which is computed using a Newton-Raphson method applied to \eqref{comp_engy_lambda}.

The post-bifurcation branch of the curve is much more nonlinear, and a linearized perturbation analysis will not capture the entire branch.
However, the linearized perturbation analysis is able to predict the post-bifurcation slope.
The stretches in \eqref{sol_bvp} are substituted with $\la_1 = \lab + \epsilon_1$ and $\la_2 = \lab + \epsilon_2$, where $\epsilon_1$ and $\epsilon_2$ are assumed to be small. 
After substitution, and elimination of higher-order terms in $\epsilon_1$ and $\epsilon_2$, \eqref{sol_bvp} becomes:
\begin{equation} \label{eq1_pert}
     (\epsilon _1 + \epsilon _2) \left(\frac{1}{\lab ^3} -2 \lab ^3 + 3 \lab ^5 \left(\ga  - \hE_3^2\right) \right) -\lab ^4-\frac{1}{\lab ^2} -3 \ga +  \lab ^6 \left(\ga  - \hE_3^2\right) =0.
\end{equation}

We further use  either of the equilibrium equations \eqref{str1} or \eqref{str1}. 
After substitution, and elimination of higher-order terms in $\epsilon_1$ and $\epsilon_2$, we get:
\begin{equation} \label{eq2_pert}
    \epsilon _1 \left(\frac{3 \ga }{\lab ^4} +\frac{3}{\lab ^6}+1 +\lab ^2 \left(\ga  -\hE_3^2 \right)\right) + 2\epsilon _2 \left(\frac{1}{\lab ^6} +  \lab ^2 \left(\ga  -\hE_3^2 \right)\right)  + \lab ^3 \left(\ga  -\hE_3^2 \right) -\frac{\ga }{\lab ^3} -\frac{1}{\lab ^5}+\lab -\that_0=0
\end{equation}
Solving \eqref{eq1_pert} and \eqref{eq2_pert} for $\epsilon_1$ and $\epsilon_2$, we can write:
\begin{align}
    \la_1 = \lab + \frac{1}{3} \left(\frac{4 \left(\lab ^7+\lab \right)}{2 \lab ^6 -3 \lab ^8 \left(\ga - \hE_3^2\right) -1}+\frac{3 \left( \left(\that_0 -2 \lab \right)\lab ^6 -2 \ga  \lab ^3\right)}{\lab ^6+ 3\ga \lab ^2 - \lab ^8\left(\ga - \hE_3^2\right) +1}+\lab \right) \\
    \la_2 = \lab + \frac{1}{3} \lab  \left(\frac{9 \ga  \lab ^2+5 \lab ^6+8}{3 \lab ^8 \left( \ga  - \hE_3^2\right) -2 \lab ^6+1}+\frac{6 \ga  \lab ^2+6 \lab ^6-3 \lab ^5 \that_0}{3 \ga  \lab ^2 - \lab ^8 \left( \ga - \hE_3^2\right) +\lab ^6+1}-2\right).
\end{align}
The equations above are linear in the load, but they provide an approximation of the post-bifurcation behavior near the bifurcation point, as shown in Figure \ref{fig:bifurcation-approx}. 

\section{Concluding Remarks}

There are two key failure modes for a dielectric elastomer: (i) dielectric breakdown, which occurs when the applied electrical voltage exceeds a critical threshold, and (ii) the triggering of mechanical instability. While the latter can sometimes be exploited for innovative designs, it is often undesirable. In this work, we derive closed-form solutions to the bifurcation problem governing the electromechanical loading of a dielectric disk. Specifically, we discover that in the presence of a through-thickness electric field, there exists a critical value beyond which an infinite mechanical force would be required to trigger Treloar-Kearsley instability. This allows us to determine the conditions necessary to design a dielectric elastomer configuration that is impervious to instability. In principle, our approach can be extended to any structural configuration, although numerical computations may be required for more complex shapes and boundary conditions.

The dielectric breakdown of DEs depends on the elastic modulus as well as the relative permittivity, defined as $(1 + \chi)$ as used in the electrical energy density in \eqref{eq:elec_energy} \cite{STARK1955Electric}. 
A high-performance DE capable of large strain under electric field reported in \cite{Li2022Dielectric} has a normalized dielectric breakdown strength $\hE_{\text{b}} = 0.6 \sqrt{3}$, obtained from experimental results. 
This breakdown strength, which is typical of high-performance DE, is well in excess of the $\hE_3$ values needed to suppress the T-K instability following the strategy proposed in this paper.

\paragraph*{Acknowledgments.}
    We thank NSF (DMS 2108784, DMREF 1921857), BSF (2018183), and AFOSR (MURI FA9550-18-1-0095) for financial support; the TCS Presidential Fellowship to Daniel Katusele for additional support; NSF for XSEDE computing resources provided by Pittsburgh Supercomputing Center; and Timothy Breitzman and Matthew Grasinger for useful discussions.


\end{document}